\documentclass[prl,reprint,nofootinbib]{revtex4-1}
\usepackage{amsmath,amssymb}
\usepackage{color,cancel}
\usepackage[all,cmtip]{xy}

\renewcommand{\1}{\it 1}

\newcommand{\3}{\it 3}

\newcommand{\C}{{\cal C}}
\newcommand{\G}{{\cal G}}
\renewcommand{\d}{{d}}

\newcommand{\ls}{\ell_{\rm s}}

\newcommand{\nn}{\nonumber}
\newcommand{\be}{\begin{equation}}
\newcommand{\ee}{\end{equation}} 

\newcommand{\blue}{\color{blue}}

\begin{document}

\title{Twelve-dimensional Effective Action and $T$-duality}

\author{Kang-Sin Choi}
\email{kangsin@ewha.ac.kr}
\affiliation{Scranton Honors Program, Ewha Womans University, Seoul 120-750, Korea}

\begin{abstract}
We propose a twelve-dimensional supergravity action, which describes low energy dynamics of F-theory. Dimensional reduction leads the theory to  eleven-dimensional, IIA, and IIB supergravities. Self-duality of the four-form field in IIB supergravity is understood. It is necessary to abandon twelve-dimensional Poincar\'e symmetry by making one dimension compact, which is to be decompactified in some region of parameter space, such that the physical degrees of freedom are the same as those of eleven-dimensional supergravity. This makes $T$-duality explicit as a relation between different compactification schemes.
\end{abstract}
\maketitle

The ideas of Kaluza and Klein (KK) \cite{KK}, generalized to higher dimensions, are beautiful ones that translate the known field degrees of freedom and their interactions into geometry of extra dimensions. 
Most of the supergravity theories, which are hoped to have intimate connection to our world, can be obtained by dimensional reduction of eleven-dimensional one \cite{Cremmer:1978km}. However, it does not directly give type IIB supergravity in ten-dimension, although their relations is well-understood in the context of string theory.

Eleven-dimensional supergravity is a low-energy description of the M-theory \cite{Horava:1995qa}. It is also shown that type IIB superstring theory is obtained by reduction of F-theory on a torus, with its complex structure identified by axion-dilaton, and the latter is shown to be $T$-dual to M-theory \cite{Vafa:1996xn}. Thus, the effective field theory of F-theory should be twelve-dimensional, however it is not easy to write down the action. One crucial difficulty might be that the twelve-dimensional minimal fermion with Lorentzian signature $(11,1)$, which must be the case for F-theory, should have superpartner components with spin higher than two in the four dimensional language, whose interacting theory would be inconsistent \cite{Nahm:1977tg}. Another obstacle is, if F-theory is dual to M-theory, there should be no surplus field degrees of freedom, although the former is a higher dimensional theory.

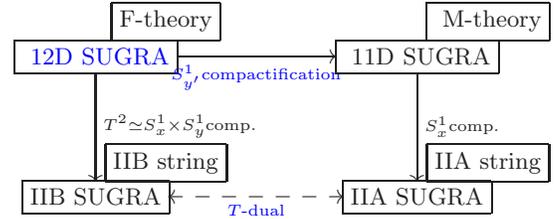
\begin{figure}[t]
$$
\xy
\xymatrixcolsep{6.3pc}
\xymatrixrowsep{3.3pc}
\xymatrix{
  *+[F]{\txt{F-theory }} &*+[F]{\txt{ M-theory}} \\
*+[F]{\txt{IIB string}}  & *+[F]{\txt{IIA string}}
}
\POS-(11.5,4.7)
\xymatrixcolsep{5pc}
\xymatrixrowsep{3.4pc}
\xymatrix{
  *+[F]{\txt{ \blue 12D SUGRA}}  \ar[r]_{\blue S^1_{y'} \text{compactification}} \ar[d]^{T^2 \simeq S^1_x \times S^1_{y}\text{comp.}} 
 &*+[F]{\txt{  11D SUGRA}} \ar[d]^{S^1_{x} \text{comp.}} \\
*+[F]{\txt{IIB SUGRA}} \ar@{<-->}[r]_{\text{\blue $T$-dual}}  & *+[F]{\txt{IIA SUGRA}}
}
\endxy
$$
\caption{Relation among superstrings and supergravities (SUGRA). In twelve-dimension, we make $T$-duality explicit in terms of compactification, by taking the other routes. The diagonal direction is the zero size limit.} \label{f:relations}
\end{figure}

An important hint comes from a careful look at the derivation of F-theory \cite{Vafa:1996xn,Denef:2008wq}. Although it is $T$-dual to M-theory, F-theory has one more dimension than the latter. Now, this extra dimension is a {\em dual} dimension to one of the dimension shared by the two theories. In other words, F-theory has {\em two redundant} dimensions which are $T$-dual to each other. Although we cannot maintain twelve-dimensional Poincar\'e symmetry fully, each ten- and eleven-dimensional theories can be symmetric in its own.  There is no contradiction if we cannot see both at once. Therefore, it is natural to keep {\em both the dimensions}. In this picture, M-theory looks like a compactification F-theory on a circle, as schematically shown in Figure \ref{f:relations}.

In this Letter, we propose the bosonic part a desired twelve-dimensional effective action, whose dimensional reductions leads to those of all known supergravities in eleven and ten dimensions, found in standard textbooks \cite{P}. Since we follow and make use of the duality relation between M- and F-theory from the eleven-dimensional supergravity, this theory shall provide the effective field theory for F-theory.

Supergravity is powerful enough in the sense that many of new results here, like existence of three-brane and generalized $T$-duality are obtained {\em without} referring to string theory. Of course, the effective field description of F-theory is timely in realistic model building, because we have so far borrowed descriptions, for instance of the gauge fields, from M-theory \cite{Beasley:2008dc,Grimm:2010ks,Choi:2013hua}.

\section{The bosonic action of twelve dimensional supergravity}

We start with the fundamental bosonic degrees of freedom of eleven-dimensional supergravity: graviton $G_{mn}$ and rank three antisymmetric tensor field $C_{map}$. The last one is promoted to a four-form field
\begin{equation}  \label{fourformpotential}
  C_{mnp}(x^m) \to {\cal C}_{mnpy'}(x^m,y'),
\end{equation}
with total antisymmetrization, for instance $\C_{mny'p} \equiv -\C_{mnpy'}$. Here $y'$ denotes the twelfth direction.
Although this field is twelve-dimensional, we do not introduce any more degrees of freedom if one of the indices is forced to be on $y'$ and the others are eleven-dimensional. The graviton is also regarded as a part of the twelve-dimensional one
\be \label{metric12D}
 ds^2 = G_{mn} dx^m dx^n + r^2 dy^{\prime 2}.
\ee

We suggest a formally twelve-dimensional action 
\begin{equation} \label{twelveDaction}
 S = \frac1{2 \kappa_{12}^2} \int \left( {\cal R}\, {*1}-\frac12 \G_{\it 5} \wedge *\G_{\it 5}   + \frac16 {\cal C}_{\it 4} \wedge  G_{\it 4} \wedge G_{\it 4} \right),
\end{equation}
with the twelve-dimensional Hodge star operator. The Ricci scalar $\cal R$ is made of the twelve-dimensional metric (\ref{metric12D}). We will define $\kappa_{12}$ shortly. 
The presence of last term is noticed in Refs. \cite{Ferrara:1996wv, Donagi:2008kj}. Other definitions and derived relations are in order.
\begin{align} 
{\cal C}_{\it4}   &= \frac{1}{3!} {\cal C}_{mnpy'} \d x^m \wedge \d x^n \wedge \d x^p \wedge  \d y' \label{C4def} \\
  &\equiv C_{\it 3} \wedge r \d y' , \nonumber \\
\G_{\it 5}& \equiv 
\d C_{\it 3} \wedge r \d y' \equiv G_{\it 4} \wedge  r \d y' .\label{G5}
\end{align}
It is important to note that the indices assume only eleven-dimensional coordinates. Therefore the action (\ref{twelveDaction}) has at best {\em eleven-}dimensional Poincar\'e invariance. Nevertheless this form is useful, since we may also have ten-dimensional invariance in which we {\em include} $y'$ and exclude some of the other directions. 
There is another loop correction term, having the form $\C_{\it 4} \wedge I_8$ where $I_8$ is again dependent on eleven-dimensional metric only, given in Ref  \cite{Duff:1995wd}.

The equation of motion and the Bianchi identity of $\C_{\it 4}$ follow
\begin{equation} \label{EOM}
 \d G_{\it 4} = 0,\quad \d {* \G_{\it 5}} = - \frac12 G_{\it 4} \wedge G_{\it 4}.
\end{equation}
Exchanging the role of the two, we also have a dual field strength 
\begin{equation} \label{duality}
* \G_{\it 5}  \equiv \d {C}_{\it 6} - \frac12 C_{\it 3} \wedge G_{\it 4},
\end{equation}
which defines a six-form ${C}_{\it 6}$. In components, the dual field strength to $\G_5$ is defined as
\be \label{dualcomp}
 (*\G_{\it 5})_{lmnpqrs} = \frac{1}{4!} \sqrt{-G} {\epsilon_{lmnpqrs}}^{tuvw y'} {\G}_{tuvw y'} ,
\ee
where the indices are raised by the metric (\ref{metric12D}). 
Note that we have converted the eleven-dimensional field $C_{\it 3}$ to the twelve-dimensional field $\C_{\it 4}=r C_{\it 3}$ using the metric (\ref{metric12D}). They should not be treated as independent degrees of freedom, otherwise we cannot match the equation of motion with the eleven-dimensional one. 

The four form structure (\ref{fourformpotential}) suggests that there is a coupled three-brane wrapped on $y'$ direction, becoming M2-brane of M-theory \cite{Choi14-2}. When a dimension is compact, this wrapping behavior should not be strange, since in the decompactification limit it becomes D3-brane along the $y$-direction, which we are familiar with. 

We consider in this letter only the bosonic degrees of freedom. The fermonic part will be dealt with elsewhere \cite{Choi14-2}.

\section{Reduction to eleven-dimensional supergravity} 

The action (\ref{twelveDaction}) is meaningful only if we take the $y'$-direction as a circle with a radius $2 \pi r$, measured in a length unit $\ell$. Dimensional reduction gives us the KK tower of the fields $\C_{\it 4}, \G_{mn}, r$ with masses 
\be \label{KKmass}
 M_k^2 = k^2 \ell^{-2} \langle r \rangle^{-2}.
\ee
All of them shall play important role later in decompactification, but we keep the zero modes only for the moment.
We can show that the kinetic terms of graviton and and three-form field become the standard form of eleven-dimensional supergravity. The last term in (\ref{twelveDaction}) is
$$
 \int {\cal C}_{\it 4} \wedge G_{\it 4} \wedge G_{\it 4} = - \int_{S^1} r \d y' \wedge \int_{M^{10,1}} C_{\it 3} \wedge G_{\it 4} \wedge G_{\it 4}.
$$
The eleven-dimensional coupling $\kappa_{11}$ may reversely define the coupling $\kappa_{12}$
\begin{equation} \label{kappa12}
  \frac{2 \pi \ell \langle r \rangle } {2\kappa_{12}^2} = \frac{1}{2 \kappa_{11}^2},
\end{equation} 
with the scale $r$ is to be fixed shortly.

\section{Reduction to IIB supergravity}

Next, we compactify two more dimensions on a torus. It has a complex structure $\tau = \tau_1 + i \tau _2$, and we take the coordinate $x,y$ such that we identify $x + \tau y  \sim x + \tau y + 2 \pi \ell \sim x + \tau y  + 2 \pi \tau \ell$. Still we keep the $y$-direction orthogonal to the other directions, as in (\ref{metric12D}). The most general metric is
\begin{equation} \label{themetric}
\begin{split}
 \d s^2 &=  L^2 \left(\d x + \tau_1 \d y  +( a_{\mu}- \tau_1 b_{ \mu} )dx^\mu  \right)^2 
                   \\
                    +& L^2 \tau_2^2 \left( \d y -  b_{\mu } \d x^\mu\right)^2 
                       +r^2 \d y^{\prime 2} +  g_{\mu \nu}' dx^\mu dx^\nu.
\end{split}
\end{equation}
From now on, fields and their Greek indices are nine-dimensional.
Here, $\{a_{\mu},\tau_1\},b_{\mu}$ are ten and nine-dimensional Lorentz vectors promoting the $S^1$ isometries of $x$- and $y$-directions, respectively, to $U(1)$ gauge symmetries. 

\begin{table}
\begin{center}
 \begin{tabular}{cccc}
 \hline \hline
10D field & type & (9+1)D components & 12D components\\ \hline
$A_{\it 1}$ & RR & $\{ A_{\mu}, A_{y}\}$ & $\{ a_\mu,\tau_1\}$ \\
$A_{\it 3}$ & RR & $\{ A_{\mu \nu \rho}, A_{\mu \nu y}\}$ & $ \{r^{-1}\C_{\mu \nu \rho y' }, r^{-1}\C_{\mu \nu y y'}  \}$\\
$B_{\it 2}$ & NSNS & $\{B_{\nu \mu },B_{\mu y} \}$ & $\{r^{-1}\C_{\mu \nu x  y'}, r^{-1}{\C}_{ \mu x y y' } \}$ \\
$b_{\it 1}$ & KK & $b_{\mu} $ & $b_\mu$ \\
\hline
 $A_{\it 4}$ & RR & $ A_{\mu \nu \rho y'}  $ & $r^{-1}{\C}_{\mu \nu \rho y' }$ \\
   $A_{\it 2}$ &RR & $\{A_{\mu \nu}, A_{\mu y'} = -A_{y' \mu} \}$ & $\{r^{-1}{\C}_{\mu \nu y y'},a_{ \mu}\}$ \\
 $A_{\it 0}$ & RR & $A$ & $\tau_1$ \\
 $B_{\it 2}$ &  NSNS & $\{B_{\nu \mu},B_{\mu y'} = - B_{y' \mu}  \}$ & $\{r^{-1}{\C}_{\mu \nu x y' },b_{\mu}\}$ \\
 $K_{\it 1}$ & KK  & $K_{\mu} $ & $r^{-1}{\C}_{\mu x y y' }$ \\
\hline 
\end{tabular} 
\end{center}
\caption{Identification of ten-dimensional fields. Indices are nine-directional and $y'$ denotes the twelfth direction. Componentwise $\C_{mnpy'}=r C_{mnp}$ as in (\ref{C4def}). After decompactifying $y'$ or $y$ directions ten-dimensional Poincar\'e covariance is recovered. } \label{t:fields}
\end{table} 

We identify the fields of IIB supergravity as in Table \ref{t:fields}. They have either all indices nine dimensional or one component fixed to be $y'$. Consider the reduction from $\G_{\alpha \beta \gamma y y'}$ to $H_{\alpha \beta \gamma} = 3 \partial_{[\alpha} B_{\beta \gamma]}$, given in (\ref{H3}) in the appendix. Neglecting the normalization, there are two possible expressions 
\begin{align}
 H_{\alpha \beta \gamma} +3 b_{[\alpha} H_{\beta \gamma]} &= H_{\alpha \beta \gamma} + 6 K_{[\alpha} \partial_{\beta} b_{\gamma]} 
\end{align}
up to a total derivative which is gauge transformation.
The left-hand side is the result of dimensional reduction of the ten-dimensional IIA field $\{ H^{(10)}_{\mu \nu \rho}, H^{(10)}_{\mu \nu} \}$ coupled to the KK field $b_\mu$, whereas the right-hand side looks as {\em dimensional reduction} of the IIB field $\{ H^{(10)}_{\mu \nu \rho}, (d b)^{(10)}_{\mu \nu } \}$ coupled to the KK field $K_\mu = r^{-1} {\C}_{\mu x y y'}$  under the metric \cite{Bergshoeff:1995as}
\begin{equation} \label{IIBmetric}
\begin{split}
 \d s^2_{10} =   r^2 (\d y' + K_\mu dx^\mu)^{2} + g_{\mu \nu}' dx^\mu dx^\nu.
\end{split}
\end{equation}
In the latter picture, the vectors $a_{\mu}$ and $b_{\mu}$ become components  $A_{\mu y'}$ and $B_{\mu y'}$, respectively, of rank two Neveu--Schwarz Neveu--Schwarz (NSNS) and Ramond--Ramond (RR) tensors.
We already have the KK tower of the fields (\ref{KKmass}) completing the fields $B_{\mu\nu}, b_\mu,K_\mu,g'_{\mu\nu}, r$ to be ten-dimensional.
This is crucial necessary condition to recover ten-dimensional Poincar\'e symmetry and fully covariant interactions. In low-energy theory, this is a possible way to see the presence of extra dimensions, if we admit that the gravitational interactions are not observable.

The RR four-form is obtained as
\begin{equation} \label{F4withyprime}
 A_{\mu \nu \rho y'} \equiv r^{-1} {\cal C}_{\mu \nu \rho y'}, \quad F_{\mu \nu \rho \sigma y'} \equiv r^{-1} {\G}_{\mu \nu \rho \sigma y'}
\end{equation}
with one of the indices fixed to be $y'$. We can perform dimensional reduction, as in (\ref{G5exp}) in the appendix (with a different nine-dimensional metric), and decompactification in the $y'$ direction with the help of one-form $K_{\it 1}$ as above. The corresponding part of the second term in (\ref{twelveDaction}) gives the kinetic term for 
\begin{equation} \label{F5}
 \textstyle  F_{\it 5}^{(10)} - \frac12 A_{\it 2}^{(10)} \wedge H_{\it 3}^{(10)}+ \frac12 B_{\it 2 }^{(10)} \wedge F_{\it 3}^{(10)} .
\end{equation}
Remember that one of the indices is {\em fixed to be $y'$} for every term in (\ref{F5}). To avoid confusion later, we name this as $ \tilde F_{\it 5}^{\rm w(10)}.$

Due to the fixing of the component in (\ref{F5}) we do not have complete ten-dimensional four-form. The other part may come from another twelve-dimensional field (\ref{duality}). The only possible nine-dimensionally covariant four-form can be\footnote{We are using the standard antisymmetric tensor notation \cite{MTW}.}
\begin{equation} \label{C4fromC6}
 A_{\mu \nu \rho \sigma} = {C}_{[\mu \nu \rho \sigma] x y}, \quad  F_{\mu \nu \rho \sigma \tau}  \equiv (d A)_{\mu \nu \rho \sigma \tau}.
\end{equation}
Due to the {\em twelve-}dimensional structure in (\ref{dualcomp}), the fields in (\ref{C4fromC6}) cannot have any index on $y'$. The left-hand side of the duality relation (\ref{dualcomp}) becomes the $dC_{\it 6}-\frac12 C_3 \wedge G_4$ as in (\ref{duality}), with two components fixed to be $x$ and $y$. By expansion and decompactification, we obtain precisely the same expression as (\ref{F5}), up to normalization, with {\em all the indices to be nine-}dimensional. Hence we may call the result as $L^{-2} \tau_2^{-1} \tilde F_{\it 5}^{\rm wo(10)}.$

Expressing the duality relation (\ref{dualcomp}) in a local Lorentz frame by ten-dimensional fields, we have
\be \label{SDdefinition}
 \tilde F_{\it 5}^{\rm wo (10)} L^{-2} \tau_2^{-1} = r {*_{10} \tilde F}^{\rm w (10)}_{\it 5},
\ee
with the ten-dimensional Hodge operation $*_{10}$.
The different components of $F_{\it 5}$ have the different origins, therefore the Lorentz symmetry is not trivial. 
For the covariance we need the same coefficient
\begin{equation} \label{r}
 r = L^{-2} \tau_2^{-1}.
\end{equation}
This means that the three radii in the $x,y,y'$ directions are inverse among themselves, so there is no point in the moduli space where we can have all the twelve dimensions nonimpact.

Therefore we have arrived at the self-duality condition for the fully covariant ten-dimensional four-form field via its modified field strength $\tilde F_{\it 5}^{(10)}$. This is mere re-expressing the relation (\ref{SDdefinition})
\begin{equation} \label{SDcondition}
 \begin{split}
  \tilde F_{\it 5}^{(10)} \equiv \tilde F_{\it 5}^{\rm w(10)} + \tilde F_{\it 5}^{\rm wo(10)} = {*_{10} \tilde F}_{\it 5}^{(10)} .
   \end{split} 
 \end{equation}
We  emphasize that this self-duality condition (\ref{SDcondition}) is the {\em defining} relation of half the components of the four-form field in (\ref{C4fromC6}).

The ten-dimensional Einstein--Hilbert term is obtained as
\be \label{Rreduction}
 \int_{T^2}  {\cal R} {*1} =(2 \pi \ell)^2\sqrt{-G'} r^{-1} R_{(10)}  d^9x \wedge dy' + \dots  
\ee
where $G' = g' r^2$ is the determinant of ten dimensional metric (\ref{IIBmetric}), with which the Ricci scalar $R_{(10)}$ is calculated. Noting that $\tau_2= g_{\rm IIB}^{-1}$, if we require $L$ should be absent from the the IIB supergravity action. Careful investigation shows that rescaling $g_{\mu \nu}' \equiv L^{-1} g_{\mu \nu}$ and $g_{y'y'} \equiv L g'_{y'y'} =L^3 \tau_2$ can pull out the overall factor $r$, which should be absorbed by the coupling
\begin{equation} \label{IIBcoupling}
\frac{1}{2 \kappa_{\rm IIB}^2} =  \frac{(2 \pi \ell  )^2 \langle r \rangle  }{2 \kappa_{12}^2} = \frac{2 \pi \ell}{2 \kappa_{11}^2}.
\end{equation}
The rescaling should also rescale the coordinate periodicity as
\be \label{stringlength}
  \ell \to  L^{-1/2} \ell \equiv \ell_{\rm s}.
\ee

Finally, dimensional reduction of the last term in (\ref{twelveDaction}) gives
\begin{equation}
 \label{IIBCS}
 \frac{1}{2 \kappa_{\rm IIB}^2}  \int F^{(10)}_{\it 5} \wedge B_{\it 2 } \wedge F_{\it 3} = \frac{1}{2 \kappa_{\rm IIB}^2}  \int \tilde F^{\rm w (10)}_{\it 5} \wedge B_{\it 2 }^{(10)} \wedge F_{\it 3}^{(10)},
\end{equation}
again with one index fixed to be $y'$ for $F_{\it 5}^{(10)}$ and $\tilde F_{\it 5}^{(10)}$. For the equality, we used the relation $F_{\3} \wedge F_{\3}=0$ and performed integration by parts. Again $\tilde F_{\it 5}^{\rm w (10)}$ can be exchanged by the covariant one (\ref{SDcondition}).  
The remaining expansions give the kinetic terms for the IIB supergravity action in the standard form \cite{Choi14-2,P}.

\section{Reduction to IIA supergravity}

We may decompactify the $y$-direction in (\ref{themetric}) using the KK field $b_\mu$. Decompactification takes place in the same way. For example, the relation (\ref{ExptoF3}) in the appendix, after the decompactification, gives the reduction of $G_{mnpq}$ to ten-dimensional fields
\be \label{F4reduction}
 L^2 (F_{\it 4} - A_{\it 1}  \wedge H_{\it 3}), 
\ee
with one of the indices fixed on $y$, whereas Eq. (\ref{G5exp}) provides the remaining components. The $A_{\1}$ is again the KK gauge field decompactifying $x$-direction. 
This gives IIA supergravity.  We identify ten-dimensional couplings $\langle L\rangle ^3 = g_{\rm IIA}^2 \equiv \langle e^{2 \Phi} \rangle$. It is straightforward to have the type IIA supergravity action, because we know it is also obtained by further compactification of eleven-dimensional supergravity action along the $x$-circle.

In the unit (\ref{stringlength}) we can naturally convert between IIA and IIB theories in ten dimensions. The relation between the two radii from (\ref{themetric}) now becomes the familiar $T$-duality relation 
\be
 R_{y}  = L^{3/2} \tau_2 \ls, \quad R_{y'} = L^{-3/2} \tau_2^{-1} \ls = \ls^2 /R_{y}.
\ee

Without referring to string theory, we can perform $T$-duality by two different compactifications, as in Figure \ref{f:relations}. In particular, the relation (\ref{r}) also allows us to interpret the KK tower of fields above Eq. (\ref{KKmass}) as ones arisen by {\em wrapping} M2-branes on the torus, whose mass is proportional to the dimensionless volume of the torus $L^2  \tau_2$ \cite{Bergshoeff:1995as,Aspinwall:1995fw,Schwarz:1983qr} . 
This will also be useful in describing physics around the self-dual radius where the two theories are not so much distinct, or in a strong coupling regime of one theory.

\subsection*{Acknowledgements}
This work is partly supported by the National Research Foundation of Korea with grant number 2012-R1A1A1040695.

\appendix
\section{Appendix}

We briefly summary technical details of dimensional reduction taking into account the metric. We have tensors in components in a local Lorentz frame, after the rescaling (\ref{stringlength}):
\begin{align}
 \C_{\alpha \beta \gamma y'} &= L^{3/2}( A_{[\alpha \beta \gamma]y'} - 3 a_{[\alpha} B_{\beta \gamma]} + 3 b_{[\alpha} A_{\beta \gamma]} \nn \\
  - & 6 a_{[\alpha} b_{\beta} K_{\gamma]}) , \\
 \C_{\alpha \beta x y'} &=   B_{\alpha \beta} + 2b_{[\alpha} K_{\beta]}, \\
 \C_{\alpha \beta y y' } &=  \tau_2^{-1} (A_{\alpha \beta} - \tau_1 B_{\alpha \beta} 
 + 2a_{[\alpha} K_{\beta]}  - 2 \tau_1 b_{[\alpha} K_{\beta]} ),\\
 \C_{\alpha x y y'} &=   L^{-3/2} \tau_2^{-1} K_\alpha.
 \label{C6reduction} 
 \end{align}
For convenience we have fixed some of the coordinates.
We have the corresponding field strengths
\begin{align}
 \G_{\alpha \beta \gamma\delta y'} & = L^2(F_{[\alpha \beta \gamma \delta] y'} - 4 a_{ [\alpha} H_{\beta \gamma \delta]} \nn \\
+ &  4 b_{ [\alpha} F_{\beta \gamma \delta]}+ 12 a_{[\alpha} b_\beta H_{\gamma \delta]}),  \label{G5exp} \\
  \G_{\alpha \beta \gamma x y'} & =  L^{1/2}( H_{\alpha \beta \gamma} +3 b_{[\alpha} H_{\beta \gamma]})
  \textstyle 
  , \label{H3} \\
 \G_{\alpha \beta \gamma y y' } & = L^{1/2} \tau_2^{-1}(F_{\alpha \beta \gamma} - \tau_1 H_{\alpha \beta \gamma}+3 a_{[\alpha} H_{\beta \gamma]} \nn \\
 -& \textstyle 3 \tau_1 b_{[\alpha} H_{\beta \gamma]})   ,\label{ExptoF3} \\
  \G_{\alpha \beta x y y'} &  \textstyle 
  =L^{-1} \tau_2^{-1}  H_{\alpha \beta} \label{H2}  . 
   \end{align}
Here $H_{\alpha \beta} \equiv 2 \partial_{[\alpha} K_{\beta]}$ and the derivative operator acts all the fields on the right.

The above relations are nine dimensional and are to be lifted to ten dimensional ones. For instance, we may rewrite the right-hand side of (\ref{G5exp}) in differential forms 
\be \label{Ftilde}
 L^2( F_{\it 5} \textstyle - \frac12 \left( A_{\it 2} \wedge H_{\it 3} -B_{\it 2 } \wedge F_{\it 3} \\
 + A_{\it 3} \wedge H_{\it 2}-K_{\it 1} \wedge F_{\it 4} \right)).
\end{equation}
This is a {\em nine}-dimensional relation with one component fixed to be on $y'$. After decompactification, we have ten-dimensional relation
\be \textstyle
L^{1/2} \tau_2^{-1} ( F^{(10)}_{\it 5} - \frac12 A_{\it 2}^{(10)} \wedge H_{\it 3}^{(10)} +\frac12 B_{\it 2 }^{(10)} \wedge F_{\it 3}^{(10)}),
\end{equation}
with the $y'$-component still fixed. There were two changes: The overall normalization has changed by the metric factor $L^{-3/2} \tau_2^{-1}$ from the $y'$-dependence, and the couplings with $K_1$ completed the covariant ten dimensional fields.

\end{document}